# Tumor suppressor and anti-inflammatory protein: an expanded view on insulin-degrading enzyme (IDE)

**Razvan Tudor Radulescu**

*Molecular Concepts Research (MCR)*

*Muenster, Germany*

E-mail: ratura@gmx.net

**Abstract.** In 1994, I conjectured that insulin-degrading enzyme (IDE) acts as an inhibitor of malignant transformation by degrading insulin and thus preventing this major growth-stimulatory hormone from binding and thereby inactivating the retinoblastoma tumor suppressor protein (RB). Ten years later, I discovered that a carboxyterminal RB amino acid sequence resembles the catalytic center of IDE. This structural homology raised the possibility that insulin degradation is a basic mechanism for tumor suppression shared by RB and IDE. Subsequently, a first immunohistochemical study on the differential expression of human IDE in normal tissues, primary tumors and their corresponding lymph node metastases further corroborated the initial conjecture on IDE being an antineoplastic molecule. In this report, it is shown that IDE harbors ankyrin repeat-like amino acid sequences through which it might bind and, as a result, antagonize the pro-inflammatory factor NF-kappaB as well as cyclin-dependent kinases (CDKs). As equally revealed here, IDE also contains 2 RXL cyclin-binding motifs which could contribute to its presumed inhibition of CDKs. These new findings suggest that IDE is potentially able to suppress both inflammation and oncogenesis by several mechanisms that ultimately ensure RB function.

Even after almost a century of intensive research, the details of insulin action and their relevance to major disease processes are still being elucidated. Towards a better understanding of this key molecule, investigations of its physical interactions with the insulin receptor, insulin-degrading enzyme (IDE) and retinoblastoma protein (RB) continue to provide important clues.

Approximately 2 years after the initial description (1) of a potential complex formation between insulin and RB in 1992 which had suggested that insulin may directly inactivate (nuclear) RB similar to the effect of RB-binding viral oncoproteins, I inferred on the basis of the previously shown cytosolic interference of IDE with insulin´s nuclear translocation (2) and a partial structural homology between IDE and RB (3) that IDE represents a tumor suppressor preventing RB inactivation (3).

A decade later, I detected in RB an amino acid sequence highly related to the catalytic center of IDE (4). This indicated a previously unknown potential for RB to degrade insulin and, moreover, significantly added to the validity of the initial IDE tumor suppressor hypothesis (3), thus collectively yielding the notion of insulin degradation possibly being a fundamental mechanism of tumor suppression (4).

A subsequent experimental study on human breast cancer tissues further reinforced the concept of a growth-inhibitory IDE protein as its presence was demonstrated to be increased (in the sense of a host-defensive phenomenon) from non-malignant to neoplastic cells pertaining to primary tumors, yet decreased from primary tumors to their corresponding lymph node metastases (5).

As it has been known for a long time that there is often a progression from (chronic) inflammation of tissues to their malignant transformation, I have now addressed the question as to whether insulin, IDE and RB may collectively be involved in this transition. A first hint in support of this anticipated relationship can be derived from the well-established association between (hyperinsulinemic) insulin resistance and inflammation (6). Such correlation implies that, as in other hyperinsulinemic states such as obesity and other potential premalignancies as well as during the aging process (7), intracellular insulin levels are probably increased in inflammatory states and this feature could be relevant to their pathogenesis. Accordingly, there is presumably also an IDE dysfunction under such conditions.

In order to substantiate this assumption on an anti-inflammatory activity of IDE, I have searched for structural similarities between IDE and I-kappaB alpha, the principal inhibitor of the pro-inflammatory protein NF-kappaB in the cytoplasm. I thereby focused primarily on those I-kappaB alpha domains mediating its binding to NF-kappaB, specifically I-kappaB alpha´s ankyrin repeats, particularly the first such repeat that is of crucial importance in this binding (8).

Interestingly, I have identified 4 ankyrin motif-like sequences in IDE (Fig. 1) that bear resemblance to ankyrin repeats in I-kappaB alpha and to alike amino acid patterns in the cyclin-dependent kinase (CDK) inhibitor p16 in the latter of which they play a key role in this protein´s binding to CDKs and its ensuing prevention of their phosphorylation and inactivation of RB (9). These findings suggest that IDE is potentially able to bind both NF-kappaB and CDKs by means of its putative ankyrin repeats and, as a result of these cytosolic encounters, inhibit their (nuclear) activity. Notably, this conclusion is supported by the fact that such amino acid signature enables I-kappaB alpha not only to block NF-kappaB, but also CDKs (10) and, conversely, p16 to abrogate CDK activity and NF-kappaB action (11).

| Sequence | Annotation |
|---|---|
| DGFTP*L*H*LA*AQQGHLEIVKL*LL*ERGADVNAQDK | human ANK consensus (9) |
| PSADW*L*AT*A*AARGRVEEVRA*LL*EAGALPNAPNS | human p16$_{11-43}$ ANK1 |
| YGRRP*I*QV*M*MMGSARVAELL*LL*HGAEPNCADPA | human p16$_{44-76}$ ANK2 |
| DGDSF*L*H*LA*IIHEEKALTME*VI*RQVKGDLAFLN | human I-kappaB alpha$_{73-105}$ ANK1 |
| LQQTP*L*H*LA*VITNQPEIAEA*LL*GAGCDPELRDF | human I-kappaB alpha$_{110-142}$ ANK2 |
| RLAWL*L*H*PA*LPSTFRSVLGAR*L*PPPERLCGFQK | human IDE$_{4-36}$ pot. ANK1 |
| REYRG*L*E*LA*NGIKVLLMSDPTTDKSSAALDVHI | human IDE$_{62-94}$ pot. ANK2 |
| VSHEH*L*E*GA*LDRFAQFFLCP*L*F*DESCKDREVNA | human IDE$_{153-185}$ pot. ANK3 |
| WRLFQ*L*E*KA*TGNPKHPFSKFGTGNKYTLETRPN | human IDE$_{199-231}$ pot. ANK4 |

**Fig. 1** Alignment of human ankyrin repeat consensus motif with similar domains in p16, I-kappaB alpha and IDE whereby 4 IDE regions are juxtaposed with 2 exemplary ankyrin signatures (from a total of more such patterns) in each of the former 2 proteins. Bold and underlined letters indicate highly conserved residues. Amino acid sequences are shown in one-letter code. "ANK" stands for ankyrin repeat motif and "pot." is the abbreviation for potential.

Moreover, I have identified (Fig. 2) that, intriguingly, IDE harbors 2 RXL cyclin-binding amino acid motifs that are similar to those present in the growth-suppressive RB family member p107 and the key CDK inhibitor p21 (12,13).

| Sequence | Annotation |
|---|---|
| K **R** R **L** F G | human p107$_{524-529}$ |
| C **R** R **L** F G | human p21$_{18-23}$ |
| K **R** R **L** I F | human p21$_{154-159}$ |
| L **R** L **L** M T | human IDE$_{686-691}$ |
| I **R** R **L** D K | human IDE$_{891-896}$ |

**Fig. 2** Alignment of RXL amino acid motifs in the human proteins p107, p21 and IDE. Crucial residues are highlighted in bold letters. Amino acid sequences are displayed in one-letter code.

Beyond its 2 RXL motifs, IDE possesses a VXL motif within its second potential ankyrin repeat motif (Fig. 1). Interestingly, such VXL motifs have previously been shown to be comparable to RXL motifs in yielding CDK inhibition (14).

Furthermore, the fact that IDE contains not only such RXL/VXL motifs, but also the above-described putative ankyrin repeats hints at this protein being a p16/p21-like CDK inhibitor rather than a substrate for CDKs.

Along with potentially serving as domains for contacting CDKs and NF-kappaB, the candidate ankyrin motifs in IDE may equally contribute to a proper conformation of the IDE catalytic center (spanning IDE residues 108-112) for the proteolytic inactivation of insulin, conceivably such that the putative IDE ankyrin repeats 1 and 2 form one flank while repeats 3 and 4 constitute the other flank of IDE´s enzymatic binding cleft for its substrate insulin.

Consequently, IDE binding of NF-kappaB is likely insufficient to counteract inflammation in those cases associated with abundant intracellular insulin as the latter may escape its degradation due to NF-kappaB´s masking of IDE´s catalytic center. As a result, insulin could move to the nucleus to physically disturb RB´s antiproliferative activity. Alternatively or additionally, insulin-bound, saturated IDE would be (sterically) unable to neutralize NF-kappaB.

In this context, it is noteworthy that the previously predicted (15) IDE contact sites for the multidrug resistance protein P-glycoprotein (P-gp), i.e. $IDE_{24-27}$ and $IDE_{84-87}$, are located within the potential IDE ankyrin repeats 1 and 2, respectively (Fig. 1). This indicates that P-gp may engender (anti-cancer) drug resistance not only through releasing insulin (15), but also NF-kappaB from IDE. This latter displacement could result from a direct competition between P-gp and NF-kappaB for the same IDE binding sites given that NF-kappaB harbors several 4-amino acid sequences (GADL, GDGL and GASL) in its p50 subunit as well as the signature GALL in its p65 subunit whereby all these tetrapeptides are highly similar to one of the anticipated (15) P-gp recognition sites for IDE, specifically the GAGL segment of P-gp (15).

Taken together, these novel findings point to a fascinating scenario in which IDE, besides degrading insulin in the cytosol, blocks NF-kappaB- hence preventing this factor´s activation of cyclin D/CDK complexes and their ensuing inactivation of RB (16)- as well as directly interferes with these complexes, thus protecting RB in multiple ways. Essentially, this report suggests that IDE- by controlling insulin-RB heterodimer-induced acceleration of cell growth, as previously outlined (17), and CDK-driven RB hyperphosphorylation, as proposed here- might represent a paradigm for uniting 2 distinct RB-defensive principles, i.e. insulin antagonism and CDK inhibition (18), in a single molecule with the potential to prevent both inflammation and cancer.